\documentclass[preprint, prX]{revtex4}

\usepackage{amsmath}    
\usepackage{graphicx}   
\usepackage{verbatim}   
\usepackage{color}      
\usepackage{subfigure}  
\usepackage{hyperref}   
\usepackage{amssymb}    
\usepackage{epsfig}
\usepackage{graphics,graphicx}
\usepackage{setspace}
\usepackage{url}
\usepackage{algorithm,algorithmic}
\usepackage{hyperref}
\newcommand{\lsem}{[\![}
\newcommand{\rsem}{]\!]}

\floatname{algorithm}{Program}

\begin{document}

\begin{abstract}
EasyTime is a domain-specific language (DSL) for measuring time during sports competitions.
A distinguishing feature of DSLs is that they are much more amenable to change, and
EasyTime is no exception in this regard. This paper introduces two new EasyTime features:
classifications of competitors into categories, and the inclusion of competitions where the number of laps must be dynamically determined. It shows how such extensions can be incrementally added into the base-language reusing
most of the language specifications. Two case studies are presented showing the suitability
of this approach.

\textit{To cite paper as follows: I. Jr. Fister, T. Kosar, I. Fister, M. Mernik, EasyTime++: A case study of incremental domain-specific language development. Information technology and control, 42(1), 77--85, 2013.
}

\end{abstract}

\title{EasyTime++: A case study of incremental domain-specific language development}

\author{Iztok Fister Jr.}
\altaffiliation{University of Maribor, Faculty of electrical engineering and computer science
Smetanova 17, 2000 Maribor}
\email{iztok.fister@guest.arnes.si}

\author{Toma\v{z} Kosar}
\altaffiliation{University of Maribor, Faculty of electrical engineering and computer science
Smetanova 17, 2000 Maribor}
\email{tomaz.kosar@uni-mb.si}

\author{Iztok Fister}
\altaffiliation{University of Maribor, Faculty of electrical engineering and computer science
Smetanova 17, 2000 Maribor}
\email{iztok.fister@uni-mb.si}

\author{Marjan Mernik}
\altaffiliation{University of Maribor, Faculty of electrical engineering and computer science
Smetanova 17, 2000 Maribor}
\email{marjan.mernik@uni-mb.si}

\maketitle

\section{Introduction}\label{sectionIntroduction}

Domain-specific languages (DSLs)  are languages tailored to specific
application domain \cite{1,2,3,4}.
They offer substantial gains regarding expressiveness and ease of use compared with
general-purpose languages (GPLs) within their domain of application \cite{5,6,7}. However, DSLs are more amenable to changes \cite{1,8}
since stakeholders' requirements frequently change. In order to design and implement
DSLs more easily, we need to develop fully modular, extensible,
and reusable language descriptions, whilst some of the descriptions could even be inferred from DSL programs \cite{9,10}.
The language designer wants to include new language features incrementally as the programming language evolves.
Ideally, a language designer would like to build a language simply by reusing different language definition
modules (language components), such as modules for expressions, declarations, etc., as well as to extend previous
language specifications. In the case of general software development the use of object-oriented techniques and concepts
like encapsulation and inheritance, greatly improves incremental software development, whilst reusability is even further
enhanced using aspect-oriented techniques \cite{11}. The object-oriented, as well as the aspect-oriented techniques and concepts,
have also been integrated into programming language specifications \cite{12,13}  making
new features more easily implemented. One of such tools, where object-oriented and aspect-oriented concepts have been incorporated,
is the LISA tool \cite{8,14}. This
paper shows how LISA is used within the incremental development of EasyTime DSL, which has been developed recently for measuring time
at different sports competitions (e.g., triathlon, cycling) \cite{15,16}. EasyTime DSL has already proved to be successful when used at real sport events
(e.g., World Championship in the double ultra triathlon in 2009, National (Slovenian) Championship in the time-trials for
cycles in 2010), so the requirements are changing quickly.
Recent extensions to EasyTime have included the possibility of classifying competitors into different categories,
where the number of laps is different for each category, and the inclusion of competitions where the number of laps can be dynamically determined during
a competition (e.g., biathlon, where the number of extra laps depends on missed shots). The objective of this paper is
to introduce EasyTime++ DSL, which supports these new extensions, as well as to show how such an extension can be incrementally developed using the introduced LISA tool.

The structure of this paper is as follows: in Section~\ref{sectionRelatedWork} an overview of the language composition is presented.
Section~\ref{sectionEasyTime} briefly introduces EasyTime DSL, whilst the core of this
paper is Section~\ref{sectionExtensions}, which describes how the extensions in EasyTime++ have been specified and
implemented. Some examples are presented in Section~\ref{sectionExamples}. The paper is concluded with
Section~\ref{sectionConclusion}, where a brief overview and word about future work is described.

\section{Related work}\label{sectionRelatedWork}
Several kinds of language composition have been identified in the literature \cite{8,17,18,19,20,21,22,23,24}. In their recent paper \cite{17}, Erdweg et al., point out that language composition has obtained
little attention, that it is still insufficiently understood, and that the terminology is confusing thus indicating
that the research is inadequate, as yet. Erdweg et al. identified the language composeability not as a property of languages themselves, but as a property
of language definition (e.g., how language specifications can be composed together).
The following types of language composition have been distinguished in \cite{17}:
language extension (which subsumes also language restriction), language unification, self-extension, and
extension composition.
In \textbf{language extension} the specifications of base language $B$ are extended with a new
language specification fragment $E$, which typically makes little sense when regarded independently from the
base language $B$. Hence, language $B$ is a dominant language, which can be a DSL or a GPL, and serves as a base for other languages.
Language extension, as a kind of language composition, is denoted as $B \triangleleft E$ indicating that the base language $B$
has been extended with the language $E$. The LISA tool supports
language extensions when single attribute grammar inheritance \cite{8} is employed. As is shown in
Section~\ref{sectionExtensions}, EasyTime++ is a language extension over the
base language EasyTime (EasyTime $\triangleleft$ EasyTime++).
In \textbf{language unification} the composition of language specifications is not based on the dominance of one language, but is
based on equal terms. The dominance of one language over another does not exist and both language specifications are
complete and standalone (note that in the case of language extension the language specifications for the extended part
makes little sense alone). Language unification, as a kind of language composition, is denoted as $L_1 \uplus_g L_2$, describing
the language composition of languages $L_1$ and $L_2$ using a glue code $g$. Since LISA supports multiple attribute grammar inheritance \cite{8},
language unification is easily achieved by inheriting both language specifications (from $L_1$ and $L_2$), where the glue code is specified
as a new language specification fragment.
In \textbf{self-extension} the language specifications do not change. The language itself is powerful enough for new extensions to be implemented
using macros, function composition, and libraries that provide domain-specific constructs. This form of language composition
is called 'pure language embedding' \cite{3}. Functional languages are these languages particularly suitable for self-extension.
Self-extension, as a kind of language composition, is denoted as $H \leftarrow E$ indicating that the host language $H$ has been self-extended
with the embedded language $E$.
The last form of language composition is \textbf{extension composition}, which describes how language specifications also support the combination of various
language compositions. That is showing how different compositions can work together. This kind of language composition can also be described as
high-order language composition. Language unification allows for such higher-order composition per se (e.g., $L_1 \uplus_g (L_2 \uplus_h L_3)$).
Whilst some other useful examples of higher-order language composition like $(B \triangleleft E_1) \triangleleft E_2$,
and $B \triangleleft (L_1 \uplus_g L_2)$ can not always be easily achieved. Extension compositions involving language extension and
language unification can also be easily achieved in the LISA tool.

\begin{figure*}[htb]  
    \begin{center}
        \includegraphics [scale=0.5]{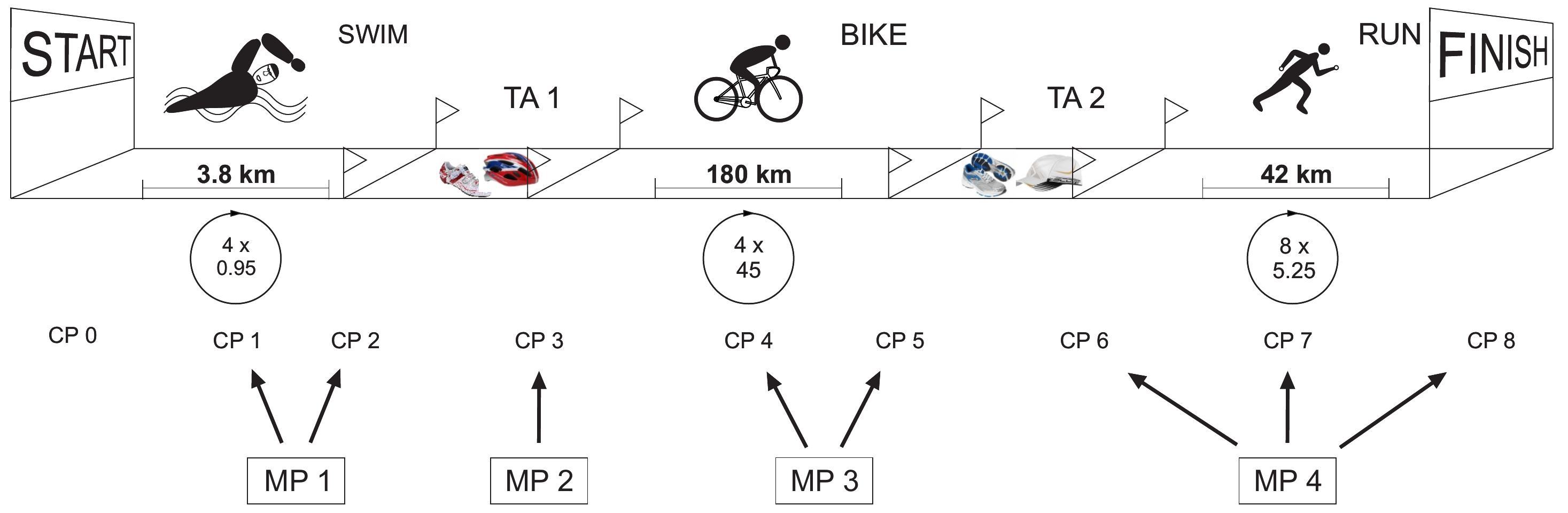}  %
        \caption{Measuring time in Ironman triathlon.}
        \label{pic:triathlon}
    \end{center}
\end{figure*}

In addition to LISA, which has been in existence since 1999, there are also other similar tools
(e.g., Phobos~\cite{18}, JastAdd~\cite{19}, Silver~\cite{20}, XMF~\cite{21}, Tatoo~\cite{22}, MontiCore~\cite{23}, JAYCO~\cite{25}, UUAG \cite{26}) that enable various language compositions. Note, that the most well-known tools for syntax and the semantic specification of programming languages, Lex and Yacc \cite{27}, don't support language composition per se. For example, language extension is possible by manually changing base language specifications $B$ by invasively adding the specification for extended language $E$. Hence, change is done in a non-disciplined manner, thus prohibiting further reuse of specifications. On the other hand, language composition can be done on top of Lex and Yacc (e.g., \cite{28}). Here, it is desirable to briefly mention JastAdd \cite{19} and Silver \cite{20}, since both are based on Attribute Grammars, as in the LISA case. JastAdd \cite{19} is centered around object-oriented representation of the abstract syntax tree (AST). Non-terminals act as abstract super classes and productions act as specialized concrete
subclasses that specify the syntactic structure, attributes, and semantic rules. All
these elements can be inherited, specialized, and overridden within subclasses. The
idea of aspect-orientation in JastAdd is to define each aspect of the language in a
separate class and then weave them together at appropriate places. The JastAdd
system is a class weaver: it reads all the JastAdd modules and weaves the fields
and methods into the appropriate classes during the generation of the AST classes.
Developers have the possibility of combining various language specifications following
the separation of different language aspects amongst different classes.
Silver \cite{20} uses a concept called 'forwarding' to achieve modular language extensions, where
the extension construct is translated into semantically equivalent constructs within the host language. Hence, forwarding only allows those new constructs that can be expressed as a combination of existing language constructs. Additional Silver features like: with-clause, auto-copying  of inherited attributes, collection attributes, pattern matching, and type-safe polymorphic lists, allow for the host language to be extended in a more flexible manner, although still restrictive.

\section{EasyTime}\label{sectionEasyTime}

EasyTime was developed for measuring time during Double ultra triathlon in 2009. At that time, the organizers of this competition were confronted with the problem of how to measure the times of competitors within three disciplines using a limited number of measuring devices. Besides this limitation, measures needed to be reliable and accurate, especially, because of its long duration. 
Although the measuring time for the triathlon was our first specific task, the goal was to develop a DSL for measuring time for any competition. A domain analysis was performed \cite{15} using feature diagrams \cite{29} with the aim of identifying common and variable concepts, their relations, and structure of particular concepts. In the case of EasyTime, the concept race consists of sub-concepts: events (e.g., swimming, cycling, and running), control points (starting and finishing lines, the number of laps), the measuring time (updating time and decrementing laps), optional transition area (difference between the finish and start times), and agents (automatic or manual). In the next step, these concepts were mapped to the context-free grammar non-terminals of EasyTime. Finally, its whole syntax and semantics were developed \cite{15}.

In order to illustrate the power of EasyTime, lets describe the Ironman triathlon, as presented in Figure~\ref{pic:triathlon}. This triathlon consists of: 3.8 km swim, 180 km cycling, and a 42 km run. These disciplines run one after another with two interruptions: In the first, those competitors who have finished with swimming prepare themselves for the cycling, whilst in the second, those competitors who have finished the cycling prepare themselves for running. Both interruptions occur within so-called transition areas. Their times spent within these areas are added to their swimming, cycling, and running times, in order to obtain the total times of specific competitors.

Typically, the organizers divide those courses on which they run particular disciplines into laps because of easier management. As can be seen in Figure~\ref{pic:triathlon}, competitors need to accomplish 4 laps of swimming, 4 laps of cycling, and 8 laps of running. These laps represent another demand for the measuring time in such competitions because, besides the intermediate times of each lap, decrementing also needs to be performed. In order to reduce the number of measuring devices, a measuring point (MP in Figure~\ref{pic:triathlon}) at which the intermediate time is measured and the number of laps is decremented, can be incorporated. In other words, when the number of laps is zero the last intermediate time becomes the final time of a specific discipline.

This characteristic of the triathlon is put to profitable use by EasyTime. In fact, EasyTime is a DSL that enables the organizers of sporting competitions to adapt measuring systems for various kinds of competitions, reduce the number of measuring devices, and achieve accuracy and reliability. The EasyTime program runs on a measuring system and employs a set of agents that control the measuring devices. For measuring time during Inronman, as illustrated in Figure~\ref{pic:triathlon}, the EasyTime program presented in Program~\ref{alg:prog} is used.

\begin{algorithm}[htb]
\caption{EasyTime program for measuring time in Ironman}
\label{alg:prog}
\scriptsize
\begin{algorithmic}[1]
\STATE 1 manual "man.dat";\ \ \ \ \ \ \ \ \ \ //\ definition of manual agent
\STATE 2 auto 192.168.225.100;\ \ \ \ // definition of automatic agent
\STATE // definition of variables
\STATE var ROUND1 := 4;
\STATE var INTER1 := 0;
\STATE var SWIM := 0;
\STATE var TRANS1 :=0;
\STATE var ROUND2 := 4;
\STATE var INTER2 :=0;
\STATE var BIKE := 0;
\STATE var TRANS2 :=0;
\STATE var ROUND3 := 8;
\STATE var INTER3 := 0;
\STATE var RUN := 0;
\STATE // definition of measuring place 1
\STATE mp[1] $\rightarrow$ agnt[1] \{
\STATE \ \ (true) $\rightarrow$  upd INTER1;
\STATE \ \ (true) $\rightarrow$ dec ROUND1;
\STATE \ \ (ROUND1 == 0) $\rightarrow$ upd SWIM;
\STATE \}
\STATE // definition of measuring place 2
\STATE mp[2] $\rightarrow$ agnt[1] \{
\STATE \ \ (true) $\rightarrow$  upd TRANS1;
\STATE \}
\STATE // definition of measuring place 3
\STATE mp[3] $\rightarrow$  agnt[2] \{
\STATE \ \ (true) $\rightarrow$  upd INTER2;
\STATE \ \ (true) $\rightarrow$  dec ROUND2;
\STATE \ \ (ROUND2 == 0) $\rightarrow$  upd BIKE;
\STATE \}
\STATE // definition of measuring place 4
\STATE mp[4] $\rightarrow$  agnt[2] \{
\STATE \ \ (true) $\rightarrow$  upd INTER3;
\STATE \ \ (ROUND3 == 8) $\rightarrow$  upd TRANS2;
\STATE \ \ (true) $\rightarrow$  dec ROUND3;
\STATE \ \ (ROUND3 == 0) $\rightarrow$  upd RUN;
\STATE \}
\end{algorithmic}
\normalsize
\end{algorithm}

At the start of Program~\ref{alg:prog}, two agents are defined: The former describes a measuring device on which manual measuring time is performed on a portable computer by an operator, whilst the latter denotes a measuring device that automatically tracks an event caused when a competitor crossing the measuring place, based on RFID technology~\cite{30}. Typically, the automatic measuring place is implemented as a mat that acts as an antenna having two functions: Firstly, the antenna induces a passive tag that is born by competitor. Secondly, this induced tag acts as a transmitter that transmits its identification code to a measuring device. The transmitted code is detected by the measuring device as an event. This event is transmitted to the measuring system and recorded into a database by an agent.

After the agents definition in Program~\ref{alg:prog}, a declaration of variables follows. For each measuring place, two variables are defined in general: an intermediate time INTERx and a laps counter ROUNDx. The final achievements of a competitor for specific disciplines are saved within variables SWIM, BIKE, and RUN.

The EasyTime program is completed by definitions of measuring places mp[$i$], where $i$ represents its identification number that must be defined uniquely. The measuring place represents a physical device that is connected to a measuring device. The measuring device can support more measuring places simultaneously.  Conversely to a measuring place, a control point (CP in Figure~\ref{pic:triathlon}) represents an event from a logical point of view and denotes the specific location on the course, where the referees need to track the time information about competitors. As a matter of fact, the control points in EasyTime are directly mapped into variables. The identification numbers of each agent responsible for transmitting the events is assigned to each measuring place. For example, manual agent agnt[$1$] in line 16 of Program~\ref{alg:prog} controls the first measuring place. 

Before recording the event into a database, a sequence of statements in curly brackets are interpreted on an abstract machine (AM). These statements are in forms of (\textbf{predicate}) $\rightarrow$ \textbf{operation}, where \textbf{operation} denotes a sequence of instructions that are executed when the value of \textbf{predicate} returns value \textit{true}. Typically, two instructions are employed in EasyTime++: \textbf{upd} and \textbf{dec}. The former update the value of variable in the database, whilst the latter decrements its value.
 
Although, DSLs can be implemented in vastly possible ways \cite{1}, an appropriate implementation when the end-users are not also  the programmers is, a compiler/interpreter approach \cite{31}. Hence, EasyTime was implemented using a compiler generator tool called  LISA \cite{8,14}. The LISA specifications include lexical, syntax and semantic specifications. Whilst classical regular expressions and BNF are used for the first two specifications, the third specifications are based on Attribute Grammars \cite{32}. One of the distinguishing features of a LISA compiler generator is that specifications (lexical, syntax, and semantics) can be easily reused and extended. An overall view of LISA specifications is given in Listing~\ref{tab:LISA-spec}. 

\begin{table}[htb]           
\caption{Overall view of LISA specifications}
\label{tab:LISA-spec}
\footnotesize
\begin{center}
\begin{tabular}{ | l | }
\hline
\ \ {\bf language } $L_1$ $[$ {\bf extends} $L_2$, ..., $L_N$ $]$ \{\\
\ \ \ \ {\bf lexicon} \{\\
\ \ \ \ \ \ $[[$P$]$ {\bf overrides} $|$ $[$P$]$ {\bf extends}$]$ R regular expr.\\
\ \ \ \ \ \ \ \ \vdots\\
\ \ \ \ \}\\
\ \ \ \ {\bf attributes} type A\_1, ..., A\_M\\
\ \ \ \ \ \ \vdots\\
\ \ \ {\bf rule} $[[$Y$]$ {\bf extends} $|$ $[$Y$]$ {\bf overrides}$]$ Z \{\\
\ \ \ \ \ \ X ::= X$_{11}$ X$_{12}$ ... X$_{1p}$ {\bf compute} \{\\
\ \ \ \ \ \ \ \ \ \ semantic functions \}\\
\ \ \ \ \ \ \ \ \vdots\\
\ \ \ \ \ \ $|$\\
\ \ \ \ \ \ \ \ X$_{r1}$ X$_{r2}$ ... X$_{rt}$ {\bf compute} \{\\
\ \ \ \ \ \ \ \ \ \ semantic functions \} \\
\ \ \ \ \ \ ;\\
\ \ \ \ \ \ \}\\
\ \ \ \ \ \ \vdots\\
\ \ \ \ {\bf method} $[[$N$]$ {\bf overrides} $|$ $[$N$]$ {\bf extends}$]$ M \{\\
\ \ \ \ \ \ operations on semantic domains\\
\ \ \ \ \ \ \}\\
\ \ \ \ \vdots\\
\ \ \} \\
\hline
\end{tabular}
\end{center}
\normalsize
\end{table}

\section{EasyTime++}\label{sectionExtensions}

EasyTime's formal description was introduced in \cite{15}, whilst the mapping of EasyTime's
denotational semantics into attribute grammars, as well as its implementation, are
presented in \cite{16}. Due to requested extensions of EasyTime the language has evolved into EasyTime++.
This section describes the formal specifications that were necessary for the change. Due to the
space constraints, we are unable to include complete specifications. Interested readers are
further referred to \cite{15,16}.
\begin{table*}[htb]        
\caption{Semantic domains in EasyTime++}
\label{tab:Semantic-domains}
\begin{center}
{
\begin{tabular}{ | l | }
\hline
  \textbf{Category}=$\{0,1,2,3 \ldots\}$ \\
  \textbf{Gender}=$\{female, male\}$ \\
  \textbf{Runners}=$(Id \times RFID \times LastName \times FirstName \times Gender \times Category)^{*}$ \\
  \textbf{State}=$\textbf{Var}\rightarrow ((\textbf{Category} \rightarrow \textbf{Integer}) \times \textbf{Truth-Value})$ \\
\hline
\end{tabular}
}
\end{center}
\end{table*}

The first small change was done within the semantic domain \textbf{Runners}, which represents a database of competitors (Listing~\ref{tab:Semantic-domains}). The additional components are now \textbf{Gender} and \textbf{Category}. Along with $Id$, $RFID$, $LastName$, and $FirstName$, the $Gender$ and $Category$ regarding competitors have added to the semantic domain \textbf{Runners}.
The second, and the most important change within the semantic domains is how the \textbf{State}, which is the mapping from variables to values, has been modelled (Listing~\ref{tab:Semantic-domains}). Within EasyTime, the \textbf{State} was a simple mapping: \textbf{State}=\textbf{Var}$\rightarrow$\textbf{Integer}, however in EasyTime++ an initial value of an attribute depends on categories, and a variable, called 'dynamicvar', can also be initialized during a run-time. Hence, the \textbf{State}
is now modelled as: $\textbf{State}=\textbf{Var} \rightarrow ((\textbf{Category} \rightarrow \textbf{Integer}) \times \textbf{Truth-Value})$

\begin{table*}[hbt]
\caption{Meaning of declarations in EasyTime++} \label{tab:Meaning-dec}
\begin{center}
\scalebox{1.1}
{
\begin{tabular}{ | l  l  l | }
\hline
  $\mathcal{D}:\textbf{Dec}\rightarrow\textbf{State}$ & $\rightarrow$ & \textbf{State} \\
  $\mathcal{D}\lsem \textbf{var} \ x := a \rsem s$ & = & $s[x \rightarrow ((\lambda category.a) \times false)]$ \\
  $\mathcal{D}\lsem \textbf{var} \ x := \{cat_1 \rightarrow a_1, cat_2 \rightarrow a_2\} \rsem s$ & = & $s[x \rightarrow (\{cat_1 \rightarrow a_1, cat_2 \rightarrow \ a_2\} \times false)]$ \\
  $\mathcal{D}\lsem \textbf{dynamicvar}\ x \rsem s$ & = & $s[x \rightarrow ((\lambda category.\perp) \times true)]$ \\
  $\mathcal{D}\lsem D_{1};D_{2} \rsem s$ & = & $\mathcal{D} \lsem D_{2} \rsem (\mathcal{D} \lsem D_{1} \rsem s)$\\
\hline
\end{tabular}
}
\end{center}
\end{table*}

Let us describe the \textbf{State} using a simple excerpt from EasyTime++ declarations. Three variables were declared within an EasyTime++ program (Program~\ref{alg:alg3a}). The first variable, $ROUND1$ specifies that all competitors need to complete 50 laps, hence in a database of competitors the attribute $ROUND1$ is set at $50$ for all competitors. The second variable, $ROUND2$, specifies that, in a case where a competitor belongs to $category=1$, he/she needs to complete $20$ laps, whilst a competitor within $category=2$ only needs $10$ laps. In the database of competitors, the attribute $ROUND2$ is initialized according to the category. For all competitors in the first category this attribute will be initialized to $20$, and for all competitors in the second category to $10$. The third variable, $PENALTY$, is a dynamic variable and its initial value for each competitor will be set during the run-time.

\begin{algorithm}[H]
\caption{Excerpt from EasyTime++ declarations}
\label{alg:alg3a}
\small
\begin{algorithmic}[1]
\STATE   var ROUND1 := 50;
\STATE   var ROUND2 := \{ (category==1) $\rightarrow$ 20,
\STATE \ \ \ \ \ \ \ \ \ \ \ \ \ \ \ \ \ \ \ \ \ \ \ \ \ \ \ \ (category==2) $\rightarrow$ 10 \};
\STATE     dynamicvar PENALTY;\ \ // definition of dynamic var.
\end{algorithmic}
\normalsize
\end{algorithm}

The \textbf{State} in EasyTime++ is mapping which maps variable names (e.g., $ROUND1$, $ROUND2$, $PENALTY$) into two components. The first component is itself a mapping from \textbf{Category} to \textbf{Integer} (e.g., 1$\rightarrow$20, 2$\rightarrow$10), whilst the second component indicates whether a variable is dynamic or not. To cope with this new model for variables in EasyTime++, the following LISA methods are needed (note that the mapping from \textbf{Category} to \textbf{Integer} can be implemented using a hashtable \cite{16}, see Program~\ref{alg:alg2a}).  

Since all changes in EasyTime++ are done in a declaration part the semantic function $\mathcal{D}$ (for full description of EasyTime semantic functions please see \cite{15,16}), which describes the meanings of the declarations needs to be changed accordingly (Listing~\ref{tab:Meaning-dec}). 

\begin{algorithm}[H]
\caption{Implementation of EasyTime++ State in LISA}
\label{alg:alg2a}
\small
\begin{algorithmic}[1]
\STATE method M\_Var \{
\STATE \ \ \ \ class Var \{
\STATE \ \ \ \ \ \ \ \ String name;
\STATE \ \ \ \ \ \ \ \ Hashtable values;
\STATE \ \ \ \ \ \ \ \ boolean isDynamic;
\STATE \ \ \ \ \ \ \ \ Var (String name, Hashtable values,
\STATE \ \ \ \ \ \ \ \ \ \ boolean isDynamic) \{
\STATE \ \ \ \ \ \ \ \ \ \ \ \ this.name = name;
\STATE \ \ \ \ \ \ \ \ \ \ \ \ this.values = values;
\STATE \ \ \ \ \ \ \ \ \ \ \ \ this.isDynamic = isDynamic;
\STATE \ \ \ \ \ \ \ \ \}
\STATE \ \ \ \ \ \ \ \ // Java methods are omitted
\STATE \ \ \ \ \ \ \ \ ...
\STATE \ \ \ \ \}
\STATE \}
\STATE
\STATE method VarEnvironment \{
\STATE \ \ \ \ import java.util.*;
\STATE \ \ \ \ public Hashtable put (Hashtable env, Var aVar) \{
\STATE \ \ \ \ \ \ \ \ env = (Hashtable)env.clone();
\STATE \ \ \ \ \ \ \ \ env.put(aVar.getName(), aVar);
\STATE \ \ \ \ \ \ \ \ return env;
\STATE \ \ \ \ \} // java method
\STATE \}  // Lisa method 
\end{algorithmic}
\normalsize
\end{algorithm}

Semantic function $\mathcal{D}$ maps the syntactic construct \textbf{Dec}, representing the declarations, into its meaning
$\textbf{State}$ $\rightarrow \textbf{State}$, which is a mapping from  \textbf{State} to \textbf{State}. Note, how the first component of \textbf{State} is defined in a case where the categories are unspecified (first equation in Listing~\ref{tab:Meaning-dec}), and in a case of dynamic variables (third equation in Listing~\ref{tab:Meaning-dec}). In the first equation, it is stated that variable $x$ is mapped to value $a$ regardless of category. The mapping function $\lambda category.a$ is a constant function.
The second equation states that variable $x$ is mapped to different values (e.g., $a_1$, $a_2$) according to different categories (e.g., $cat_1$, $cat_2$), whilst in the third equation, the variable $x$ is mapped to undefined value $\perp$ regardless of category. In the case of dynamic variables the second component of $(\textbf{Category} \rightarrow \textbf{Integer}) \times \textbf{Truth-Value}$ is true, otherwise it is false.

The aforementioned changes in formal specifications of EasyTime++ also require changes in the implementation part. Note that changes are required in the lexical part (new keywords category and dynamicvar, new separator), syntax part (new syntax rules for declarations), as well as in the semantic part (new semantics for declarations). All the other parts of EasyTime (e.g., agents, measuring places, statements) \cite{16} are intact and hence can be completely reused. Since EasyTime is implemented in LISA, which supports attribute grammar inheritance \cite{8}, and where lexical, syntax and semantic specifications can be inherited, it was natural to extend EasyTime specifications written in LISA for implementing EasyTime++, thus achieving incremental language development. Program~{\ref{alg:alg1a}} shows the LISA specification of EasyTime++. Note, how all EasyTime specifications have been reused ('language EasyTime++ extends EasyTime'). In the inherited specifications it was necessary to override rule $Dec$, which contained syntactic and semantic specifications for declarations, add some new grammar productions and their semantics (rule $Categories$), as well as add new attribute $varvalues$ of type $Hashtable$, which were attached to the non-terminal $CTGRS$, and extend regular definitions for $Separator$ and $Keyword$. Overall less than 70 lines of LISA specifications have been newly written to obtain the complete compiler for EasyTime++. Note that this is an example of language extension where language specifications' fragment (Program~{\ref{alg:alg1a}}) alone does not make any sense and can not exist without base-language specifications (for complete EasyTime specifications in LISA see \cite{16}). Hence, this kind of language composition can be denoted as EasyTime $\triangleleft$ EasyTime++.

\begin{algorithm}
\caption{LISA specification of EasyTime++}
\label{alg:alg1a}
\small
\begin{algorithmic}[1]
\STATE language EasyTime++ extends EasyTime \{
\STATE  lexicon  \{
\STATE \ \ \ \ extends Separator \  ,
\STATE \ \ \ \ extends Keyword category | dynamicvar
\STATE \}
\STATE attributes Hashtable *.varvalues;
\STATE \ \ \ \ rule extends Start \{
\STATE \ \ \ \ \ \ \ \ compute \{ \}
\STATE \}
\STATE  rule overrides Dec \{\\
\STATE \ \ \ \ DEC ::= var \#Id :=  \#Int ; compute \{\\
\STATE \ \ \ \ \ \ \ \ // category is not specified; isDynamic=false
\STATE \ \ \ \ \ \ \ \ DEC.outState = put(DEC.inState,
\STATE \ \ \ \ \ \ \ \ \ \ new Var(\#Id.value(),
\STATE \ \ \ \ \ \ \ \ \ \ put(new Hashtable(), "0",
\STATE \ \ \ \ \ \ \ \ \ \ Integer.valueOf(\#Int.value()).intValue()), false));
\STATE \ \ \ \ \};
\STATE \ \ \ \ DEC ::= dynamicvar \#Id ; compute \{
\STATE \ \ \ \ \ \ \ \ // category can not be specified; isDynamic=true
\STATE \ \ \ \ \ \ \ \ DEC.outState = put(DEC.inState,
\STATE \ \ \ \ \ \ \ \ \ \ new Var(\#Id.value(), null, true));
\STATE \ \ \ \ \};
\STATE \ \ \ \ DEC ::= var \#Id := \{ CTGRS \} ; compute \{
\STATE \ \ \ \ \ \ \ \ // categories are specified and can't be dynamic
\STATE \ \ \ \ \ \ \ \ DEC.outState = put(DEC.inState,
\STATE \ \ \ \ \ \ \ \ \ \ new Var(\#Id.value(), CTGRS.varvalues, false));
\STATE \ \ \ \ \};
\STATE \}
\STATE rule Categories \{
\STATE \ \ \ \ CTGRS ::= ( category == \#Int ) $\rightarrow$  \#Int , CTGRS
\STATE \ \ \ \ \ \ compute \{
\STATE \ \ \ \ \ \ \ \  CTGRS[0].varvalues = put(CTGRS[1].varvalues,
\STATE \ \ \ \ \ \ \ \ \ \ \#Int[0].value(),
\STATE \ \ \ \ \ \ \ \ \ \ Integer.valueOf(\#Int[1].value()).intValue());
\STATE \ \ \ \ \};
\STATE \ \ \ \ CTGRS ::= ( category == \#Int ) $\rightarrow$  \#Int compute \{
\STATE \ \ \ \ \ \ \ \ CTGRS.varvalues = put(new Hashtable(),
\STATE \ \ \ \ \ \ \ \ \ \ \#Int[0].value(),
\STATE \ \ \ \ \ \ \ \ \ \ Integer.valueOf(\#Int[1].value()).intValue());
\STATE \ \ \ \ \};
\STATE \}
\STATE \ ...
\STATE \ \ \ \ // LISA methods
\STATE \}
\end{algorithmic}
\normalsize
\end{algorithm}

\section{Examples}\label{sectionExamples}

In order to test EasyTime++ DSL two case-studies were performed:

\begin{itemize}
\item cyclo-cross Grand-prix, and
\item biathlon.
\end{itemize}

The former was experienced in practice, whilst the latter could be taken as proof of concept. In the rest of this section, both case-studies are discussed in detail.

\subsection{Case-study 1: Cyclo-cross Grand-Prix}

This case-study tested the introduction of categories in EasyTime++. Cyclo-cross is a relatively new sport that typically takes place in winter and is dedicated to cycle road-riders who are preparing for the new season. Races usually consist of several laps of a short course featuring pavements, wooded trails, grass, steep hills, and obstacles.

\begin{figure}[htb]  
    \begin{center}
        \includegraphics [scale=0.38]{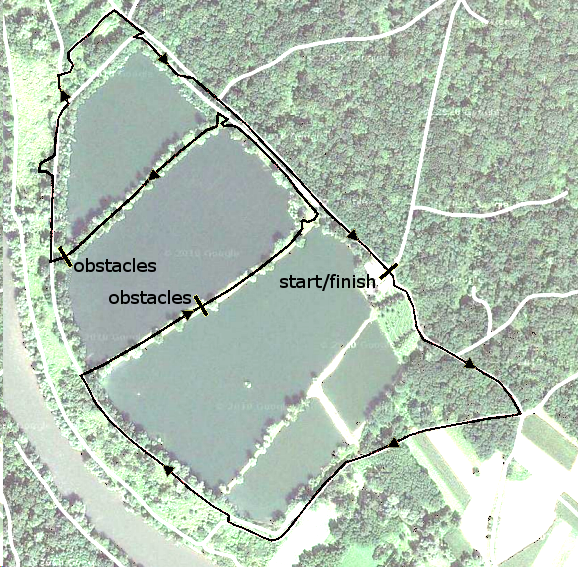}  %
        \caption{Track layout of the cyclo-cross competition.}
        \label{pic:cyclo}
    \end{center}
\end{figure}

In this case-study, one lap of 2.5 km was used (Figure~\ref{pic:cyclo}). According to the number of laps, the competitors were divided into three categories, as follows:

\begin{itemize}
\item 4 laps: junior men and women up to 15 years old (U-15),
\item 6 laps: junior men and women up to 19 years old (U-19), and
\item 9 laps: absolute categories (U-23, Elite, Masters).
\end{itemize}

In order to make the competition more interesting, the organizers allowed all the competitors onto the course simultaneously. Only one measuring device with two measuring places was needed for measuring this competition because the course passed at one location. Here, the intermediate times of laps were measured and, thereby, decremented the laps' counters of specific competitors. When the laps counter reached zero the finish time of the competitor was reported. However, how many laps to go depended on the category to which the specific competitor belonged.

The EasyTime++ program for measuring time in this competition can be seen by Program~\ref{alg:cyclo}. Note that here both measuring places, i.e., mats, were laid so that the whole length of the finish line was captured. In line with this, a competitor can cross either of both mats. As a result, the programs for both measuring devices are the same, and work in parallel. 

\begin{algorithm}[H]
\caption{EasyTime++ program for measuring time in cyclo-cross competition}
\label{alg:cyclo}
\small
\begin{algorithmic}[1]
\STATE 2 auto 192.168.225.100;\ \ // definition of agent
\STATE var BIKE := 0;
\STATE var ROUND1 := \{ (category==1) $\rightarrow$ 4,
\STATE \ \ \ \ (category==2) $\rightarrow$ 6, (category==3) $\rightarrow$ 9 \};
\STATE // definition of measuring place 1
\STATE mp[1] $\rightarrow$ agnt[2] \{
\STATE \ \ (true) $\rightarrow$ dec ROUND1;
\STATE \ \ (ROUND1 == 0) $\rightarrow$ upd BIKE;
\STATE \}
\STATE // definition of measuring place 2
\STATE mp[2] $\rightarrow$ agnt[2] \{
\STATE \ \ (true) $\rightarrow$ dec ROUND1;
\STATE \ \ (ROUND1 == 0) $\rightarrow$ upd BIKE;
\STATE \}
\end{algorithmic}
\normalsize
\end{algorithm}

In summary, measuring time in cyclo-cross performed well with EasyTime++. Although the organizers prepared three different lengths of courses, six different lists of results were obtained according to gender. Fortunately, in our case the gender could be handled by a database system, whilst the EasyTime++ program was unaware of it.

\subsection{Case-study 2: Biathlon}

\begin{figure}[htb]  
    \begin{center}
        \includegraphics [scale=0.7]{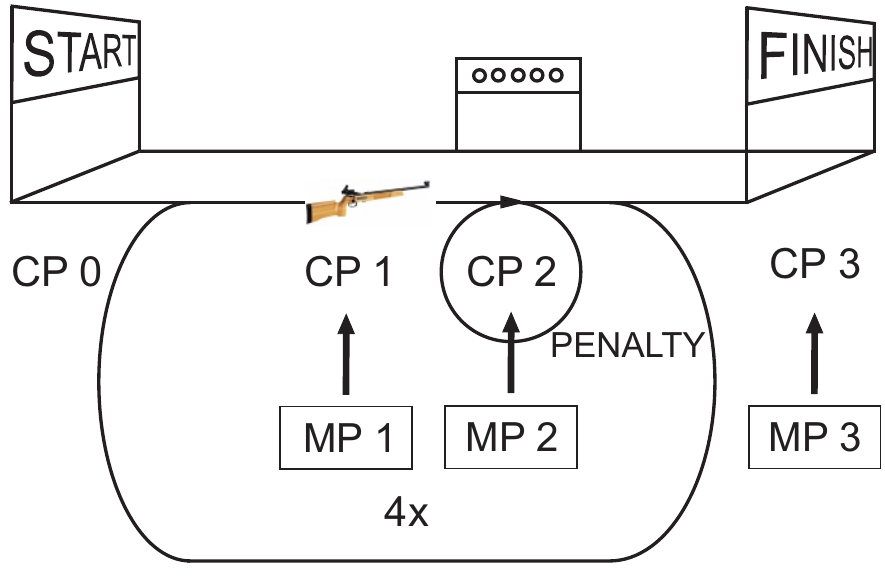}  %
        \caption{Measuring time in biathlon competitions.}
        \label{pic:biathlon}
    \end{center}
\end{figure}

A biathlon was the second case-study for EasyTime++. Biathlon refers specifically to the winter sport that combines cross-country skiing and rifle shooting. As can be seen from Figure~\ref{pic:biathlon}, competitors start with skiing. Skiing is interrupted by rifle shooting. Typically, biathlon consists of 4 laps of skiing. The shooting appears close to the end of a lap. Two positions for competitors are allowed when shooting, i.e., prone and standing. Interestingly, the number of missed shoots is penalized by the additional number of penalty laps. Note that the time spent within the penalty laps are added to the total time of the competitor. The time of the penalty lap is typically taken to be between 20-30 seconds.

The EasyTime++ program for measuring time in biathlon can be seen by Program~\ref{alg:biathlon}. Three measuring devices are needed to cover this competition. Each measuring device realizes one measuring place. Moreover, each measuring place also represents a control point. In contrast to Ironman, in a biathlon time  spent in penalty loops is of no interest in the preferred race. Here, only the total time of competitor is important.

\begin{algorithm}[htb]
\caption{EasyTime++ program for measuring time in biathlon competition}
\label{alg:biathlon}
\small
\begin{algorithmic}[1]
\STATE 1 auto 192.168.225.110;\ \ // definition of agent 1
\STATE 2 auto 192.168.225.100;\ \ // definition of agent 2
\STATE var ROUND := 4;
\STATE var RUN := 0;
\STATE dynamicvar PENALTY;\ // definition of dynam.variable
\STATE // definition of measuring place 1
\STATE mp[1] $\rightarrow$ agnt[1] \{
\STATE \ \ (true) $\rightarrow$ upd PENALTY;
\STATE \}
\STATE // definition of measuring place 2
\STATE mp[2] $\rightarrow$ agnt[2] \{
\STATE \ \ (true) $\rightarrow$ dec PENALTY;
\STATE \}
\STATE // definition of measuring place 3
\STATE mp[3] $\rightarrow$ agnt[2] \{
\STATE \ \ (true) $\rightarrow$ dec ROUND;
\STATE \ \ (ROUND == 0) $\rightarrow$ upd RUN;
\STATE \}
\end{algorithmic}
\normalsize
\end{algorithm}

In summary, the first device represents the special measuring device for counting hits. The agent assigned to this device puts the number of missed hits into the database variable PENALTY, dynamically. Note that this device is treated in EasyTime like an ordinary measuring device. The second measuring device is dealt with by counting the penalty laps, whilst the third device measures the final time.

\section{Conclusions}\label{sectionConclusion}

Easy language composition is still an open-issue within programming language research. In particular, a new young field
of software language engineering is of interest regarding engineering principles when constructing new languages, whether general-purpose or
domain-specific. A language designer would like to build a new language simply by composing different components and/or extending
previous components. This paper has presented EasyTime++ DSL as a language extension of EasyTime, where the base language specifications written in the LISA compiler
generator have been extended with new features, thus enabling the introduction of categories into competitions, and those new competitions where the 
number of laps is dynamically determined. The implemented multiple attribute grammar inheritance in LISA enables easy language composition since
lexical, syntax, and semantic specifications can be reused and extended. In such a manner, an incremental language development using LISA has been demonstrated.
The suitability of EasyTime++ was shown in two case studies: cyclo-cross Grand-prix and a biathlon.
More extensive experimental work, which would include other types of language compositions and more DSLs, is also planned in the future.

\bigskip{\small \smallskip\noindent Updated 3 March 2013.}

\begin{thebibliography}{9}

\bibitem{1} 
\textbf{M. Mernik, J. Heering, A. M. Sloane.} When and how to develop domain-specific languages. \emph{ACM Computing Surveys}, 2005, 37(4), 316--344.

\bibitem{2} 
\textbf{A. van Deursen, P. Klint, J. Visser.} Domain-specific languages: an annotated bibliography. \emph{ACM SIGPLAN Notices}, 2000, 35(6), 26--36.

\bibitem{3} 
\textbf{P. Hudak.} Building domain-specific embedded languages. \emph{ACM Computing Surveys}, 1996, 28(4es).

\bibitem{4} 
\textbf{M. Fowler.} Domain Specific Languages. \emph{Addison-Wesley Professional}, 2010.

\bibitem{5} 
\textbf{M. J. Varanda Pereira, M. Mernik, D. da Cruz, P. R. Henriques.} Program comprehension for domain-specific languages. \emph{Computer Science and Information Systems}, 2008, 5(2), 1--17.

\bibitem{6} 
\textbf{T. Kosar, N. Oliveira, M. Mernik, M. J. Varanda Pereira, M. {\v C}repin{\v s}ek, {\mbox D. da} Cruz, P. R. Henriques.} Comparing General-Purpose and Domain-Specific Languages: An Empirical Study. \emph{Computer Science and Information Systems}, 2010, 7(2), 247--264.

\bibitem{7} 
\textbf{T. Kosar, M. Mernik, J. C. Carver.} Program comprehension of domain-specific
and general-purpose languages: comparison using a family of experiments. \emph{Empirical software engineering}, 2012, 17(3), 276--304.

\bibitem{8} 
\textbf{M. Mernik, V. {\v Z}umer.} Incremental programming language development. \emph{Computer
Languages, Systems and Structures}, 2005, 31(1), 1--16.

\bibitem{9} 
\textbf{D. Hrn{\v c}i{\v c}, M. Mernik, B. R. Bryant.} Embedding DSLs into GPLs: A Grammatical Inference Approach.
\emph{Information Technology and Control}, 2011, 40(4), 307--315.

\bibitem{10} 
\textbf{D. Hrn{\v c}i{\v c}, M. Mernik, B. R. Bryant, F. Javed.} A memetic grammar inference algorithm for language learning.
\emph{Applied Soft Computing}, 2012, 12(3), 1006--1020.

\bibitem{11} 
\textbf{G. Kiczales.} Aspect-oriented programming. \emph{ACM Computing Surveys}, 1996, 28(4), Article No. 154.

\bibitem{12} 
\textbf{P. Klint, T. van der Storm, J. J. Vinju.} Term rewriting meets aspect-oriented programming. \emph{Processes, Terms and
Cycles: Steps on the Road to Infinity. Springer-Verlag}, 2005, 88--105.

\bibitem{13} 
\textbf{M. Mernik, D. Rebernak.} Aspect-Oriented Attribute Grammars. \emph{Electronics and Electrical Engineering}, 2011, 10(116), 99--104.

\bibitem{14} 
\textbf{P. R. Henriques,  M. J. Varanda Pereira, M. Mernik, M. Leni{\v c}, J. Gray, H. Wu.} Automatic generation of language-based tools using LISA. \emph{IEE Proceedings - Software Engineering}, 2005, 152(2), 54--69.

\bibitem{15} 
\textbf{I. Jr. Fister, I. Fister, M. Mernik, J. Brest.} Design and implementation of domain-specific language Easytime. \emph{Computer
Languages, Systems and Structures}, 2011, 37(4), 276--304.

\bibitem{16} 
\textbf{I. Jr. Fister, M. Mernik, I. Fister,  D. Hrn{\v c}i{\v c}.} Implementation of EasyTime Formal Semantics using a LISA Compiler Generator. \emph{Computer Science and Information Systems}, 2012, 9(3): 1019--1044.

\bibitem{17} 
\textbf{S. Erdweg, P. G. Giarrusso, T. Rendel.} Language Composition Untangled. \emph{In: Proceedings of Workshop on Language Descriptions, Tools and Applications (LDTA'12)}. \emph{Published online at http://www.informatik.uni-marburg.de/$\sim$seba/publications/languagecomposition. pdf}, 2012.

\bibitem{18} 
\textbf{A. Granicz, J. Hickey.} Phobos: A front-end approach to extensible compilers. \emph{In: Proceedings of the 36th Annual Hawaii International Conference on System Sciences (HICSS36)}, 2003.

\bibitem{19} 
\textbf{G. Hedin, E. Magnusson.} JastAdd: an aspect-oriented compiler construction system. \emph{Science of Computer Programming}, 2003, 47(1), 37--58.

\bibitem{20} 
\textbf{E. Van Wyk, D. Bodin, J. Gao, L. Krishnan.} Silver: An extensible attribute grammar system. \emph{Science of Computer Programming}, 2010, 75(1), 39--54.

\bibitem{21} 
\textbf{T. Clark, P. Sammut, J. Willans.} Superlanguages: Developing languages and applications with XMF. \emph{Published online at hhtp://bit.ly/HiTOKp}, 2008.

\bibitem{22} 
\textbf{J. Cervelle, R. Forax, G. Roussel.} A simple implementation of grammar libraries. \emph{Computer Science and Information Systems}, 2007, 4(2), 65--77.

\bibitem{23} 
\textbf{H. Krahn, B. Rumpe, S. Voelkel.} MontiCore: Modular development of textual domain specific languages. \emph{In: Proceedings of the 30th International Conference on Software Engineering (SLE 2008)}, 2008, 925--926.

\bibitem{24} 
\textbf{A. Stonis, D. Rubliauskas, J. Blonskis.} Mapping Syntax Extensions. \emph{Information Technology and Control}, 2002, 24(3), 35--48.

\bibitem{25} 
\textbf{J. Porub{\"a}n, M. Forg\'{a}\v{c}, M. Sabo, M. B\`{i}h\'{a}lek.} Annotation Based Parser Generator. \emph{Computer Science and Information Systems}, 2010, 7(2), 291--307.

\bibitem{26} 
\textbf{M. Viera, S. D. Swierstra, A. Middelkoop.} UUAG Meets AspectAG: How to make Attribute Grammars First-Class. \emph{In: Proceedings of Workshop on Language Descriptions, Tools and Applications (LDTA'12)}. \emph{Published online at http://www.cs.uu.nl/research/techreps/repo/CS-2011/2011-029.pdf}, 2012.

\bibitem{27} 
\textbf{D. Brown, J. Levine, T. Mason.} lex \& yacc. \emph{O'Reilly Media, 2nd Edition}, 1992.

\bibitem{28} 
\textbf{X. Wu, B.R. Bryant, J. Gray, M. Mernik.} Component-based LR parsing. \emph{Computer Languages, Systems and Structures}, 2010, 36(1), 16--33.

\bibitem{29} 
\textbf{V. \v{S}tuikys, R. Dama\v{s}evicius.} Measuring Complexity of Domain Models Represented by Feature Diagrams. \emph{Information Technology and Control}, 2009, 38(3), 179--187.

\bibitem{30} 
\textbf{K. Finkenzeller.} RFID Handbook. \emph{John Wiley \& Sons}, 2010.

\bibitem{31} 
\textbf{T. Kosar, P. E. Mart{\'{i}}nez L{\'{o}}pez, P. A. Barrientos, M. Mernik.} A Preliminary Study on Various Implementation Approaches of Domain-Specific Language. \emph{Information and Software Technology}, 2008, 50(5), 390--405.

\bibitem{32} 
\textbf{J. Paakki.} Attribute Grammar Paradigms - A High-Level Methodology in Language Implementation. \emph{ACM Computing Surveys}, 1995, 27(2), 196--255.

\end{thebibliography}
\end{document}